%% file: dvcs_prl.tex
\documentstyle[twocolumn,epsf,floats,aps]{revtex}

\newcommand{\pEpg}{$\vec ep~\rightarrow~ep\gamma$}
\newcommand{\Epg}{$ep~\rightarrow~ep\gamma$}
\newcommand{\Epgg}{$ep~\rightarrow~ep\gamma \gamma~$}
\newcommand{\Eppi}{$ep~\rightarrow~ep\pi^0~$}
\newcommand{\EpX}{$ep~\rightarrow~epX~$}

\begin{document}

\onecolumn
\title{{\large\bf Observation of Exclusive Deeply Virtual
Compton Scattering in Polarized Electron Beam Asymmetry Measurements}}
\input{memb}
\date{\today}
\maketitle

\wideabs{

\begin{abstract}                % DON'T CHANGE THIS LINE

We report the first results of the beam spin asymmetry measured in the
reaction \pEpg~ at a beam energy of 4.25 GeV. A large asymmetry with a
$\sin\phi$ modulation is observed, as
predicted for the interference term of Deeply Virtual Compton Scattering
and the Bethe-Heitler process. The amplitude of this modulation is
$\alpha~=~0.202\pm 0.028$. In leading-order and leading-twist pQCD,
the $\alpha$ is directly
proportional to the imaginary part of the DVCS amplitude. 

\end{abstract}

\centerline{PACS numbers: 13.60.Fz, 14.20.Dh, 24.85.+p}

}

\indent

High-energy lepton nucleon scattering experiments revealed the quark structure 
of the nucleon and were employed to measure the longitudinal quark 
momentum distributions as well as the quark spin distributions.      
The recently developed formalism \cite{Ji97,Rady,CFS,MR} 
for the QCD description of deeply 
exclusive leptoproduction reactions introduces Generalized Parton 
Distributions (GPDs). They carry new information about the dynamical
degrees of freedom inside the nucleon not accessible in inclusive
experiments. It has been shown that in the Bjorken scaling 
regime (high photon virtuality $Q^2$, high energy transfer at 
fixed Bjorken scaling variable $x_B$) the scattering amplitude for 
exclusive processes can be factorized into a hard-scattering 
part (exactly calculable in perturbative QCD) and a nucleon structure 
part parametrized via GPDs. This is called the ``handbag approximation'', 
depicted in Fig. 1a. 
In addition to the dependence on the parton momentum fraction $x$ GPDs 
depend on two more parameters, the skewedness $\xi$, and the momentum 
transfer $t$ to the baryonic system. In the
Bjorken regime $\xi \rightarrow x_B/(2-x_B)$, and is related to the momentum 
imbalance  between the struck quark and the quark 
that is put back into the final state baryon.  
The momentum transfer $t$, which can be varied independently in exclusive 
processes, measures the transverse momentum transfer between interacting 
quarks. The GPDs reduce to the longitudinal 
parton distributions in the 
limit $\xi \rightarrow 0, ~ t \rightarrow 0$. 
 
Deeply exclusive experiments probe the nucleon wave 
function at the quark-gluon level, expressed in terms of two 
spin-dependent and two spin-independent GPDs. 
Deeply Virtual Compton Scattering 
(DVCS), 
i.e. electroproduction of photons from nucleons in the Bjorken regime
is most suitable for 
studying GPDs at moderately high energies. While exclusive meson production
requires high energies and high photon virtualities to reach the Bjorken 
regime, DVCS may access the GPDs already at $Q^2$ as low 
as 1 (GeV/c)$^2$ \cite{GVG1,GVG2}. 

In this letter we present the first observation of the fully exclusive DVCS 
signal in the beam spin asymmetry measured in the reaction 
$\vec ep \rightarrow ep\gamma$. The measurement was performed using a
4.25 GeV
longitudinally polarized electron beam and the CEBAF Large Acceptance 
Spectrometer (CLAS) \cite{CLAS} at the Thomas Jefferson National Accelerator 
Facility. Observation of the beam-spin asymmetry in semi-exclusive 
production of high energy photons has been reported recently by the HERMES 
collaboration \cite{HERMES}, and cross section measurements of DVCS by the H1 
collaboration at DESY at very small $x_B$ \cite{h1}, which is complementary to 
the valence quark regime of this measurement.

%%%%%%%%%%%%%%%%%%% Figure : BH and  DVCS %%%%%%%%%%%%%%%%%%%%%%%
\begin{figure}
\vspace{40mm} 
{\includegraphics{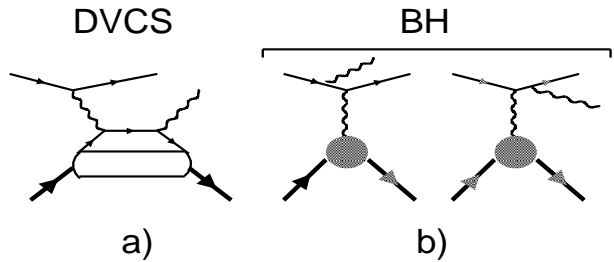}}
\caption
{Feynman diagrams for DVCS and Bethe-Heitler processes
contributing to the amplitude of \Epg~ scattering.}
\label{bhdvcs}
\end{figure}
%%%%%%%%%%%%%%%%%%%%%%%%%%%%%%%%%%%%%%%%%%%%%%%%%%%%%%%%%%%%%%%%%%%%

The experiments measures the DVCS contribution through its interference with 
the Bethe-Heitler process. In contrast to DVCS, where the photon is
emitted from the nucleon,  BH photons are emitted from the incoming or
scattered electrons (Fig.1). While the BH cross section in most of the 
kinematic region is much larger than the DVCS cross section, the interference 
between the two amplitudes boosts the effect of DVCS and produces a 
large cross section difference for electron helicities aligned or anti-aligned 
with the electron beam. In this difference, the large helicity-independent
BH contribution drops out, and only the helicity-dependent interference
term remains \cite{Die97}: 

\begin{eqnarray}
&&{d^4\sigma^+ \over {dQ^2 dx_B dt d\phi}}~-~{d^4\sigma^- \over {dQ^2
dx_B dt d\phi}} \nonumber \\
&\propto& ~Im\left( T^{DVCS} \right)T^{BH}
\propto a\cdot Im\tilde M^{1,1}\sin\phi \nonumber \\
&+&~b\cdot Im\tilde M^{0,1}\sin 2\phi
+O\left( \frac {1}{Q^2} \right)+\cdots ~.
\label{eq:csd}
\end{eqnarray}

\noindent
Here $a$ and $b$ are independent of $\phi$, the angle between the
lepton and hadron planes, and $\tilde M^{1,1}$ and $\tilde M^{0,1}$
are the helicity amplitudes for transverse and longitudinal virtual
photons, respectively. The cross
section for positive (negative) electron beam helicity
is denoted with $\sigma^+$ ($\sigma^-$). This difference measures 
directly the imaginary part of the DVCS amplitude, 
since the BH amplitude can be calculated exactly. 
In leading-order only $\tilde
M^{1,1}$ contributes \cite{Die97}.

Experimentally it is much simpler to measure an asymmetry
$A~=~(d^4\sigma^+~-~d^4\sigma^-)/(d^4\sigma^+~+~d^4\sigma^-)$
rather than a cross section difference.
This asymmetry gives the relative strength of the relevant
amplitudes and has a more complex dependence on $\phi$ than the cross
section difference, due to $\phi$ terms in the denominator. 
In the Bjorken limit, however, the $\phi$ dependence of the denominator is
expected to be small.

In this analysis, we used CLAS electroproduction data taken in
March 1999.
Scattering of 4.25 GeV longitudinally polarized electrons from
a liquid hydrogen target was studied in a wide range of kinematics.
The analyzed event sample 
corresponds to an integrated luminosity of 1.3 fb$^{-1}$.  

The reaction \Epg~ is identified by analyzing the missing mass squared
distributions in the reaction \EpX:
$M_x^2=(\nu+M-E_p)^2-(\vec q-\vec P_p)^2$,
where $\nu$ and $\vec q$ are the virtual photon energy and momentum,
$E_p$ and $\vec P_p$ are the energy and the momentum of the recoil
proton, and $M$ is the mass of the proton.

The main 
background to the single photon final state is from $\pi^0$
production. At large missing energy the missing mass 
resolution of CLAS is not sufficient to separate single photons and 
$\pi^0$s event-by-event.
The number of single photon events was determined using a
fitting technique that analyzes the line shape of the $M_x^2$ 
distributions. As a fit function the sum of two Gaussians and a third
order polynomial is used:

\begin{eqnarray}
F(M_x^2)&=&
N_\gamma \cdot e^{-(M_x^2~-~M_\gamma^2)^2/2\sigma_\gamma^2} \nonumber \\
~&+& N_{\pi^0} \cdot e^{-(M_x^2~-~M_\pi^2)^2/2\sigma_\pi^2}
\nonumber \\
~&+& \sum_{j=0}^{3} P_j\cdot (M_x^2)^j.
\label{eq:ff}
\end{eqnarray}

\begin{figure}
\vspace{70mm} 
{\includegraphics{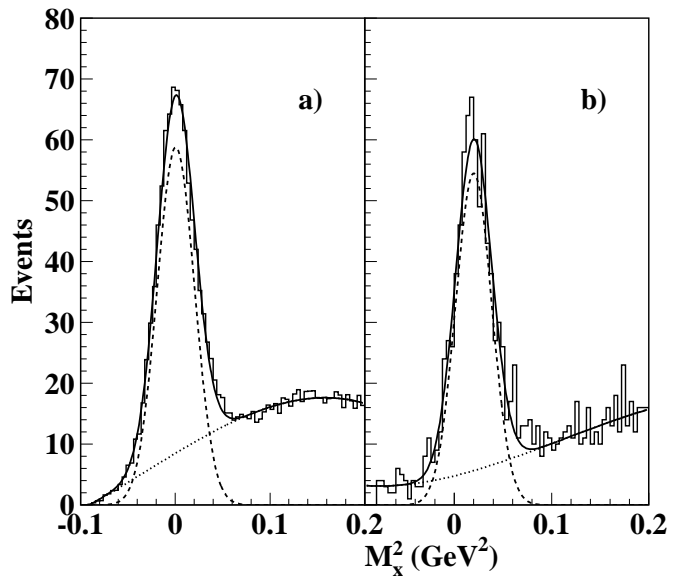}}
\caption
{Missing mass squared distribution of the detected (ep) system for
a) \Epg~ and b) \Eppi. In each plot the solid line is the fit to the sum
of a Gaussian and the third order polynomial distribution. The dashed curve
corresponds to the Gaussian function and the dotted curve represents
the polynomial function.}
\label{pig}
\end{figure}

\noindent
The Gaussians represent the $M_x^2$
distributions for single photon and single pion
final states. The
polynomial is used to model the smooth background that arises
from radiative processes. The parameters $M_\gamma^2$ and
$\sigma_\gamma^2$ are determined from the fit
to the $M_x^2$ distribution of selected Bethe-Heitler events (Figure
\ref{pig}.a). For $M_\pi^2$ and $\sigma_\pi^2$, the results of a
similar fit to the
$M_x^2$ distribution of selected events with final state
\Epgg are used (Figure \ref{pig}.b). In these events both photons are
detected in the CLAS electromagnetic calorimeter;
they have an invariant mass near the $\pi^0$ mass. 
In Figure \ref{pig} the curves represent 
the fit to a sum of the Gaussian and a third order polynomial. The
samples of BH and $\pi^0$ events were selected in the same kinematical
range as the main sample for DVCS studies.

The $M_x^2$ distributions are analyzed in eleven $\phi$ bins.
The final fit to the $M_x^2$ distributions with the function
in Eq. \ref{eq:ff} is performed separately 
for each helicity state and for the sum.
In the fit only two parameters are adjusted: the number of
single photon ($N_\gamma$) and the number of $\pi^0$ ($N_{\pi^0}$) events.
For each $\phi$ bin the shape of the background was fixed
from a fit to the points in the summed $M_x^2$ distribution 
that excluded the $M_x^2$ range from -0.07 GeV$^2$ to 0.08
GeV$^2$. The relative magnitude of the background 
for distributions at positive and negative helicities are found from
the fit to the same $M_x^2$ range as for the sum. 

%%%%%%%%%%%%%%%%%%% Figure : pion and photon MM2 %%%%%%%%%%%%%%%%%%%%%%%%%%
%%%%%%%%%%%%%%%%%%%%%%%%%%%%%%%%%%%%%%%%%%%%%%%%%%%%%%%%%%%%%%%%%%%

\begin{figure}
\vspace{70mm} 
{\includegraphics{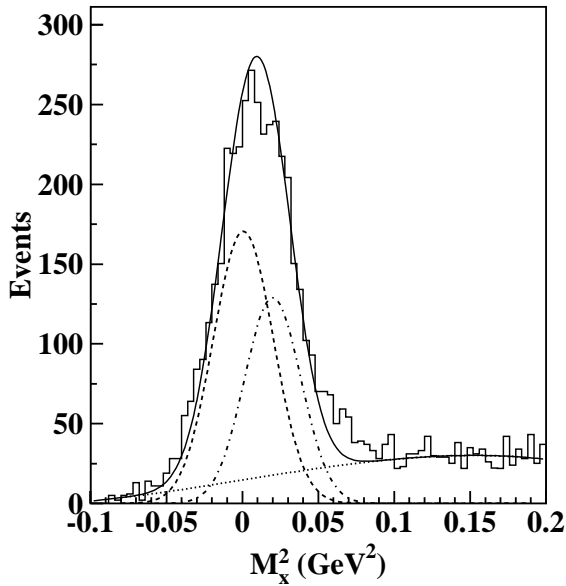}}
\caption
{Missing mass squared distribution for the reaction
\EpX. Events are integrated in the range of 
$\phi$ from 70$^\circ$ to 110$^\circ$. The curves are described in the text.}
\label{mm2phi3}
\end{figure}

Figure \ref{mm2phi3} shows a typical fit result for the sum of 
two helicities at $\phi$ = 90$^\circ$. 
The solid line corresponds to the fitted
function $F(M_x^2)$, the dashed line represents the Gaussian
for the missing photon, and the dashed-dotted line is the Gaussian 
for the missing $\pi^0$. The dotted line represents the
polynomial background. In this bin the number of photon and pion 
events are $N_\gamma=4201 \pm 109$ and 
$N_{\pi^0}=2010 \pm 101$, respectively. The $\chi^2$ per degree of freedom of
the fit is $1.8$. Similar results have been obtained for
all kinematical bins.

%%%%%%%%%%%%%%%%%%% Figure :MM2 Fit  %%%%%%%%%%%%%%%%%%%%%%%%%%%%%%%%
%%%%%%%%%%%%%%%%%%%%%%%%%%%%%%%%%%%%%%%%%%%%%%%%%%%%%%%%%%%%%%%%%%%%%%%%

The fitted number of single photon events are used to calculate
the beam spin asymmetry as:

\begin{eqnarray}
{A~=~\frac{1}{P_e} \frac {\left(N^+_\gamma~-~N^-_\gamma\right)}
{\left(N^+_\gamma~+~N^-_\gamma\right)}},
\label{eq:bsan}
\end{eqnarray}
\noindent
where $P_e$ is the beam polarization, 
$N^{+(-)}_\gamma$ is the number of 
\Epg~ events at positive (negative) beam helicity. The average beam 
polarization, $P_e =70\%$, was measured 
using M{\o}ller scattering. 

\begin{figure}
\vspace{80mm} 
\centering
{\includegraphics{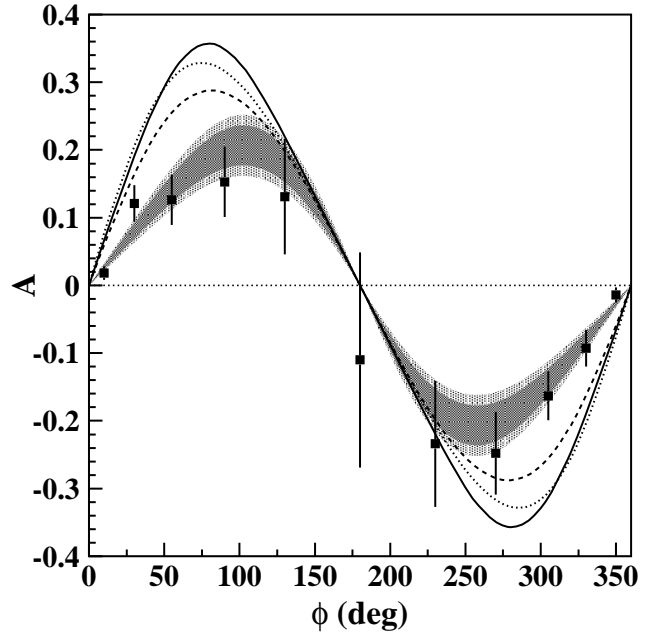}}
\caption
{$\phi$ dependence of the beam spin asymmetry A. 
The dark shaded region is the range
of the fitted function $A(\phi)$ defined by the statistical errors of 
parameters $\alpha$ and $\beta$, the light shaded region includes
systematic uncertainties added linearly to the statistical uncertainties. 
The curves are model calculations according to Refs. [6,11] and are
discussed in the text.} 
\label{asym42}
\end{figure}

In Figure \ref{asym42} we present our main result, the $\phi$ 
dependence of $A$. 
Data in each $\phi$ bin are integrated in the range of $Q^2$ from
1.00 (GeV/c)$^2$ to 1.75 (GeV/c)$^2$ and $-t$ from
0.1 (GeV/c)$^2$ to 0.3 (GeV/c)$^2$.
The error bars shown are statistical. Most of the
systematic uncertainties related to the experiment do not contribute
to $A$. Only the error in the measurement of beam
polarization, $\pm 1.65\%$, remains.  
There is also a systematic error in the
calculation of $N_\gamma$ due to the determination of the
mean and standard deviation of the Gaussian functions for the photon and pion
missing mass squared distributions, and also due to the fit procedure
to the $M_x^2$ distributions. 
These errors are defined as a deviation of $A$ from its
central value, when the mean and standard deviation of Gaussians 
are shifted within their errors, and when different fit techniques
were used (see \cite{ANOTE} for details).

The data points are fitted with the function $A\left( \phi
\right)=\alpha \sin\phi +\beta \sin 2\phi$. 
The fitted parameters are $\alpha=0.202 \pm 0.028^{stat} \pm 0.013^{sys}$ and 
$\beta=-0.024\pm 0.021^{stat} \pm 0.009^{sys}$. In the Bjorken regime
$\beta$ should vanish, leaving only the contribution from transverse photons
(see e.g. Ref.\cite {Die97}). 
In Figure \ref{asym42} the dark shaded
region corresponds to the range of the fitted function within the
statistical uncertainties of $\alpha$ and $\beta$. 
The light shaded region includes systematic uncertainties on these 
parameters, estimated using the method described above. 

%%%%%%%%%%%%%%%%%%% Figure : asymmetry %%%%%%%%%%%%%%%%%%%%%%%%%%%%%%%%
%%%%%%%%%%%%%%%%%%%%%%%%%%%%%%%%%%%%%%%%%%%%%%%%%%%%%%%%%%%%%%%%%%%%%%%%%%%%%%%%%%%%

The resulting asymmetry is in a good agreement with a $\sin\phi$
modulation.
Curves in Figure \ref{asym42} show the results of theoretical
calculations from Refs. \cite{GVG1,REV,CALC} at fixed values of $Q^2=1.25$
(GeV/c)$^2$, $x_B=0.19$, and $-t=0.19$ (GeV/c)$^2$. 
The limited experimental information does not allow to unambiguously
extract GPDs from the measurement. Description of GPDs that model the
$\xi$ and $t$ dependencies are therefore used to predict observables
accessible in experiments.
The dashed curve is a calculation
at leading-twist \cite{GVG1} and no $\xi$ dependence in the
evaluation of GPDs, the
dotted curve is leading-twist with $\xi$ 
dependence \cite{GVG1}, and the
solid curve includes twist-3 \cite{TW3,BEL01} effects. All three
calculations include the $D$-term in the parameterization of the GPDs
\cite{Pol99}, which is related to double pion
contributions. For a more detailed
description of the model assumptions we refer to a recent review 
\cite{REV}. We have estimated that the model asymmetries would be reduced
by about 7$\%$ if they are averaged over the experimental acceptances, bringing
them somewhat closer to the measured data points.

Although the experimental results are close to the lower range of 
theoretical predictions, none of the calculations is in agreement 
with our data. This could be due to several factors which need 
to be studied in future research. Firstly, parameterizations of the $\xi$ 
dependence of the GPDs were 
modeled in only a few different ways. Secondly, higher order perturbative 
QCD corrections 
have not been included. They tend to reduce the measured 
asymmetry \cite{FRMcD}.  
Thirdly, it is possible that the average $Q^2$ and $W$ of our 
measurements are not sufficiently high to fully reach the Bjorken 
regime, so that some non-perturbative corrections may be needed. 
This can be tested in a future experiment by measuring DVCS at the 
same $x_B$ and $t$ but at higher $Q^2$ and $W$. Such experiments 
have been approved and are currently in preparation \cite{PROP6,ADVCS}.

In conclusion, we have presented the first measurement of the beam
spin asymmetry in 
exclusive electroproduction of real photons in the deeply inelastic
regime. 
We see a clear asymmetry, as expected from the interference of
the DVCS and BH processes (Eq.\ref{eq:csd}).  
Our results agree in sign, and are not far in magnitude 
to predictions based on available models of GPD parameterizations. 
This strongly supports 
expectations that DVCS will allow access to GPDs at relatively low
energies and momentum transfers.
This opens up a new
avenue for the study of nucleon structure, which is inaccessible in
inclusive scattering experiments. 
Further measurements at higher beam energy are planned, which will
allow significant
expansion of the $Q^2$ and $x_B$ range covered in these
studies. The high luminosity available for these measurements will
make it possible to map out
details of the $Q^2$, $x_B$, and $t$ dependences of GPDs.

We would like to acknowledge the outstanding efforts of the staff of the
Accelerator Division and the Physics
Divisions, and the Hall B technical staff that made this experiment possible. 
We would also like to acknowledge useful discussions with
A. Radyushkin. Also many thanks to M. Vanderhaeghen and L. Moss\'e 
for help in the calculations. 

This work was supported in part by the Istituto Nazionale di Fisica 
Nucleare, the French Centre National de la Recherche Scientifique, 
the French Commissariat \`{a} l'Energie Atomique, the U.S. Department of 
Energy, the National 
Science Foundation and the Korean Science and Engineering Foundation.
The Southeastern Universities Research Association (SURA) operates the 
Thomas Jefferson National Accelerator Facility for the United States 
Department of Energy under contract DE-AC05-84ER40150.

\end{document}

%% file: memb.tex
\def\cnuva{$^{1}$}
\def\jlab{$^{2}$}
\def\asuaz{$^{3}$}
\def\ucla{$^{4}$}
\def\cmupa{$^{5}$}
\def\cuawdc{$^{6}$}
\def\cea {$^{7}$}
\def\connecticut{$^{8}$}
\def\pascal{$^{9}$}
\def\duke{$^{10}$}
\def\edinburgh{$^{11}$}
\def\fiu{$^{12}$}
\def\fsu{$^{13}$}
\def\frascati{$^{14}$}
\def\genova{$^{15}$}
\def\gwudc{$^{16}$}
\def\itep{$^{17}$}
\def\jmuva{$^{18}$}
\def\knukorea{$^{19}$}
\def\umma{$^{20}$}
\def\mit{$^{21}$}
\def\unhdurham{$^{22}$}
\def\nsuva{$^{23}$}
\def\ohio{$^{24}$}
\def\oduva{$^{25}$}
\def\ipn{$^{26}$}
\def\uppa{$^{27}$}
\def\rpi{$^{28}$}
\def\rubltx{$^{29}$}
\def\urva{$^{30}$}
\def\usc{$^{31}$}
\def\utep{$^{32}$}
\def\uvch{$^{33}$}
\def\vpsu{$^{34}$}
\def\cwm{$^{35}$}
\def\yerevan{$^{36}$}

\author{
        S.~Stepanyan,\cnuva$^,$\jlab\ 
        V.D.~Burkert,\jlab\ 
        L.~Elouadrhiri,\cnuva$^,$\jlab\  
        G.S.~Adams,\rpi\  
        E.~Anciant,\cea \
        M.~Anghinolfi,\genova\        
        B.~Asavapibhop,\umma\ 
        G.~Audit,\cea \   
        T.~Auger,\cea \ 
        H.~Avakian,\frascati\
        J.~Ball,\asuaz\
        S.~Barrow,\fsu\ 
        M.~Battaglieri,\genova\ 
        K.~Beard,\jmuva\ 
        M.~Bektasoglu,\oduva\ 
%        B.L.~Berman,\gwudc\ 
		P.~Bertin,\pascal\
        N.~Bianchi,\frascati\ 
        A.~Biselli,\rpi\ 
        S.~Boiarinov,\itep\
        B.E.~Bonner,\rubltx\ 
        S.~Bouchigny,\jlab$^,$\ipn\ 
        D.~Branford,\edinburgh\
%        W.J.~Briscoe,\gwudc\
        W.K.~Brooks,\jlab\ 
        J.R.~Calarco,\unhdurham\  
        D.S.~Carman,\ohio\
        B.~Carnahan,\cuawdc\ 
%        C.~Cetina,\gwudc\ \thanks{Present address: 
%Department of Physics, Carnegie Mellon University, Pittsburgh, PA 15213.}
        L.~Ciciani,\oduva\
        P.L.~Cole,\utep$^,$\jlab\ 
        A.~Coleman,\cwm\ 
\thanks{Present address: 
Systems Planning and Analysis, 2000 North 
Beauregard Street, Suite 400, Alexandria, VA 22311.} 
        D.~Cords,\jlab\ 
        P.~Corvisiero,\genova\ 
        D.~Crabb,\uvch\ 
        H.~Crannell,\cuawdc\
        J.~Cummings,\rpi\
        P.V.~Degtiarenko,\jlab\ 
		H.~Denizli\uppa\  
        L.C.~Dennis,\fsu\ 
        E.~De\, Sanctis,\frascati\ 
        R.~DeVita,\genova
% ...
\\*[0.2cm](The CLAS Collaboration)
}

%%%%%%%%%%%%%%%%%%%%%%%%%%%%%%%%%%%%%%%%%%%%%%%%%%%%%%%%%%%%%%%%%%%%%%%%%

\newpage

\address{
\cnuva Christopher Newport University, Newport News, VA 23606, USA\\
\jlab Thomas Jefferson National Accelerator Facility, 
Newport News, VA 23606, USA\\
\asuaz Arizona State University, Department of Physics and Astronomy, 
Tempe, AZ 85287, USA\\
\ucla University of California at Los Angeles, 
Department of Physics and Astronomy, Los Angeles, CA 90095, USA\\
\cmupa Carnegie Mellon University, Department of Physics, 
Pittsburgh, PA 15213, USA\\
\cuawdc Catholic University of America, Department of Physics, 
Washington, DC 20064, USA\\
\cea  CEA Saclay, DAPNIA-SPhN, F91191 Gif-sur-Yvette Cedex, France\\
\connecticut University of Connecticut, Physics Department, Storrs, 
CT 06269, USA\\
\pascal University Blaise Pascal/IN2P3, France\\
\duke Duke University, Physics Department, Durham, NC 27706, USA\\
\edinburgh Edinburgh University, Department of Physics and Astronomy, 
Edinburgh EH9 3JZ, United Kingdom\\
\fiu Florida International University, Department of Physics, Miami, 
FL 33199, USA\\
\fsu Florida State University, Department of Physics, Tallahassee, 
FL 32306, USA\\
\frascati Istituto Nazionale di Fisica Nucleare, 
Laboratori Nazionali di Frascati, C.P. 13, 00044 Frascati, Italy\\
\genova Istituto Nazionale di Fisica Nucleare, Sezione di Genova
e Dipartimento di Fisica dell'Universita, 16146 Genova, Italy\\
\gwudc The George Washington University, Department of Physics, 
Washington, DC 20052, USA\\
\itep Institute of Theoretical and Experimental Physics, Moscow, 117259, 
Russia\\
\jmuva James Madison University, Department of Physics, Harrisonburg, 
VA 22807, USA\\
\knukorea Kyungpook National University, Department of Physics, 
Taegu 702-701, South Korea\\
\umma University of Massachusetts, Department of Physics, Amherst, 
MA 01003, USA\\
\mit M.I.T.-Bates Linear Accelerator, Middleton, MA 01949, USA\\
\unhdurham University of New Hampshire, Department of Physics, 
Durham, NH 03824, USA\\
\nsuva Norfolk State University, Norfolk, VA 23504, USA\\
\ohio Ohio University, Department of Physics, Athens, OH 45701, USA\\
\oduva Old Dominion University, Department of Physics, Norfolk, 
VA 23529, USA\\
\ipn Institut de Physique Nucleaire d'Orsay, IN2P3, BP 1, 91406 Orsay, France\\
\uppa University of Pittsburgh, Department of Physics, 
Pittsburgh, PA 15260, USA\\
\rpi Rensselaer Polytechnic Institute, Department of Physics, 
Troy, NY 12181, USA\\
\rubltx Rice University, T.W. Bonner Nuclear Laboratory, 
Houston, TX 77005-1892, USA \\
\urva University of Richmond, Department of Physics, Richmond, VA 23173, USA\\
\usc University of South Carolina, Department of Physics, 
Columbia, SC 29208, USA\\
\utep University of Texas, Department of Physics, El Paso, TX 79968, USA\\
\uvch University of Virginia, Department of Physics, 
Charlottesville, VA 22903, USA\\
\vpsu Virginia Polytechnic Institute and State University, 
Department of Physics, Blacksburg, VA 24061, USA\\
\cwm College of William and Mary, Department of Physics, 
Williamsburg, VA 23185, USA\\
\yerevan Yerevan Physics Institute, 375036 Yerevan, Armenia\\
}